\documentclass[12pt,preprint]{aastex}

\shorttitle{Chandra Observation of MS~1054$-$0321}

\shortauthors{Jeltema et al.}

\begin{document}

\title{Chandra X-Ray Observatory Observation of the High-Redshift Cluster MS~1054$-$0321}

\author{Tesla E. Jeltema, Claude R. Canizares, Mark W. Bautz and Michael R. Malm}

\affil{Center for Space Research and Department of Physics}
\affil{Massachusetts Institute of Technology}
\affil{Room 37-241, 70 Vassar Street, Cambridge, MA 02139-4307}

\email{tesla@space.mit.edu, crc@space.mit.edu, mwb@space.mit.edu, malm@space.mit.edu}

\author{Megan Donahue}

\affil{Space Telescope Science Institute}
\affil{3700 San Martin Drive, Baltimore, MD 21218}

\email{donahue@stsci.edu}

\author{Gordon P. Garmire}
\affil{The Pennsylvania State University}
\affil{525 Davey Lab, University Park, PA 16802}

\email{gpg2@psu.edu}

\begin{abstract}

We observed MS~1054$-$0321, the highest redshift cluster of galaxies in the {\itshape Einstein\/} Medium Sensitivity Survey (EMSS), with the {\itshape Chandra\/} ACIS-S detector.  We find the X-ray temperature of the cluster to be $10.4^{+1.7}_{-1.5}$ keV, lower than, but statistically consistent with, the temperature inferred previously.  This temperature agrees well with the observed velocity dispersion and that found from weak lensing.  We are also able to make the first positive identification of an iron line in this cluster and find a value of $0.26\pm0.15$ for the abundance relative to solar, consistent with early enrichment of the ICM.  We confirm significant substructure in the form of two distinct clumps in the X-ray distribution.  The eastern clump seems to coincide with the main cluster component.  It has a temperature of $10.5^{+3.4}_{-2.1}$ keV, approximately the same as the average spectral temperature for the whole cluster.  The western clump is cooler, with a temperature of $6.7^{+1.7}_{-1.2}$ and may be a subgroup falling into the cluster.  Though the presence of substructure indicates that this cluster is not fully relaxed, cluster simulations suggest that we will underestimate the mass, and we can, therefore, use the mass to constrain $\Omega_m$.  From the overall cluster X-ray temperature we find the virial mass of the cluster to be at least $4.5 \times 10^{14} h^{-1} M_{\odot}$.  We revisit the cosmological implications of the existence of such a hot, massive cluster at a relatively early epoch.  Despite the lower temperature, we still find that the existence of this cluster constrains $\Omega_m$ to be less than one.  If $\Omega_m = 1$ and assuming Gaussian initial perturbations, we find the probability of observing MS~1054 in the EMSS is $\sim 7 \times 10^{-4}$.

\end{abstract}

\keywords{galaxies: clusters: general, individual(MS~1054$-$0321) --- X-rays: galaxies:clusters --- cosmology:observations --- dark matter}

\section{INTRODUCTION}

Clusters of galaxies are the largest and most recently virialized objects in the universe.  Massive clusters represent the extreme end of initial density perturbations and are, therefore, extremely sensitive to the density parameter $\Omega_m$.  If $\Omega_m = 1$, the number density of clusters evolves quickly with redshift, and massive clusters must have formed recently (Carlberg et al. 1997).  However, for a low density universe ($\Omega_m < 1$), the evolution with redshift is much slower, and massive clusters must form early to account for their present number density.  The existence of massive clusters at high redshift, therefore, places strong constraints on $\Omega_m$.  The virial mass of a cluster can be related to its X-ray temperature, and so $\Omega_m$ can be constrained from the evolution of cluster temperature with redshift (Oukbir \& Blanchard 1992; Donahue \& Voit 1999; Eke, Cole, \& Frenk 1996).

In this paper, we present the {\itshape Chandra\/} X-ray Observatory observation of MS1054$-$0321, the highest redshift cluster in the Einstein Extended Medium Sensitivity Survey (EMSS; Gioia et al. 1990; Henry et al. 1992) with z $= 0.83$.  This cluster has previously been observed at X-ray wavelengths with {\itshape ASCA\/} and {\itshape ROSAT\/}.  Its mass has been estimated from the {\itshape ASCA\/} X-ray temperature of $12.3^{+3.1}_{-2.2}$ keV (Donahue et al. 1998 hereafter D98), from $\beta$-model fits to the {\itshape ROSAT/HRI\/} data (Neumann \& Arnaud 2000,  hereafter NA00), and, at optical wavelengths, from its weak lensing signal (Hoekstra, Franx, \& Kuijken 2000; Luppino \& Kaiser 1997) and observed velocity dispersion (Tran et al. 1999).  All of these methods indicate that MS~1054 is a massive cluster, which in conjunction with its high redshift implies $\Omega_m < 1$.  Substructure, an indication that the cluster may not be completely relaxed, was seen in both the {\itshape ROSAT/HRI\/} observation and the weak lensing data (D98; NA00; Hoekstra et al. 2000).

Using {\itshape Chandra\/}, we make a more accurate temperature determination.  We also examine in greater detail the substructure in the X-ray distribution.  Lastly, we estimate the mass and velocity dispersion of MS~1054 and investigate the constraints this cluster places on an $\Omega_m = 1$ universe.  Throughout the paper, we use $H_0 = 100 h$ km s$^{-1}$ Mpc$^{-1}$.  For $q_0 = 0.5$ and $\Lambda = 0$, one arcminute is 249 $h^{-1}$ kpc at the cluster's redshift.

\section{OBSERVATIONS}

{\itshape Chandra\/} observed MS~1054 with the back-illuminated ACIS-S3 detector on 2000 April 21-22 for 91 ks.  To create a clean ``events file'' for analysis, we kept only {\itshape ASCA\/} grades 0, 2, 3, 4, and 6\footnote{{\itshape Chandra\/} Proposers' Observatory Guide http://asc.harvard.edu/udocs/docs/docs.html, section ``Technical Description'' subsection ``ACIS''}.  We then examined the satellite aspect and light curve to eliminate time intervals of bad aspect or high background.  The net useful exposure time was then 88 ks.  The spectral analysis was limited to the 0.8-7 keV range.  Point sources were detected using {\itshape wavdetect\/}, a wavelet source detection program in the Chandra Interactive Analysis of Observations Software (CIAO), with a significance threshold of $10^{-6}$.  Twenty-three sources were removed from the data.  For fitting, all spectra were extracted in PI (pulse height-invariant) channels, which correct for the gain difference between different regions of the CCD, and grouped to give a minimum of 20 counts per energy bin.  The background was estimated from the local background on the ACIS-S3 chip.  The net cluster count rate was 0.13 counts s$^{-1}$ in a $2^{\prime}$ radius region and in the 0.3-7.0 keV band.  Spectra were analyzed using XSPEC (v11.0.1). We generally used redistribution matrix (RMF) and ancillary response (ARF) files based on the center of a given spectrum's extraction region.  Choosing RMFs and ARFs for different points in the cluster did not significantly affect our fits.

\section{ANALYSIS}

\subsection{Cluster Properties}

In order to find the overall cluster temperature, a spectrum was extracted from a $1.5^{\prime}$ (374 $h^{-1}$ kpc) radius circular aperture surrounding the cluster.  The cluster center was taken to be R.A.(2000) = 10$^h$56$^m$58$^s$.6 and decl.(2000) =  $-$03$^{\circ}$37$^{\prime}$36$^{\prime\prime}$.7, which corresponds to the best-fit center from NA00.  The spectrum was fit with a Raymond-Smith thermal plasma model (Raymond \& Smith 1977; updated to 1992 version) with foreground absorption.  The absorption was fixed at the galactic value of $3.6 \times 10^{20}$ atoms cm$^{-2}$ (Dickey \& Lockman 1990), the redshift was fixed at 0.83, and the iron abundance and temperature were free to vary.  The best-fit temperature is $10.4^{+1.7}_{-1.5}$ keV with an abundance of $0.26\pm0.15$ relative to the abundances of Anders \& Grevessa (1989).  The quoted uncertainties are 90\% confidence levels for the two free parameters.  MS~1054 has bolometric L$_{x} = 1.2 \times 10^{45}$ h$^{-2}$ ergs s$^{-1}$ and a luminosity in the 2-10 keV band of $6.3 \times 10^{44}$ h$^{-2}$ ergs s$^{-1}$ ($q_0 = 0.1$).  The detection of the iron emission line allows us to fit for the redshift, which gives $z = 0.83\pm0.03$.  Using a mekal model instead of a Raymond-Smith did not affect these results, and the intrinsic cluster absorption was consistent with zero. 

Figure 1 shows the binned spectrum and best fit folded model.  The fit is good with a reduced $\chi^2$ of 1.03 for 239 degrees of freedom.  Figure 2 shows the $\chi^2$ contours for the iron abundance versus cluster temperature.

\begin{figure}
\epsscale{0.6}
\plotone{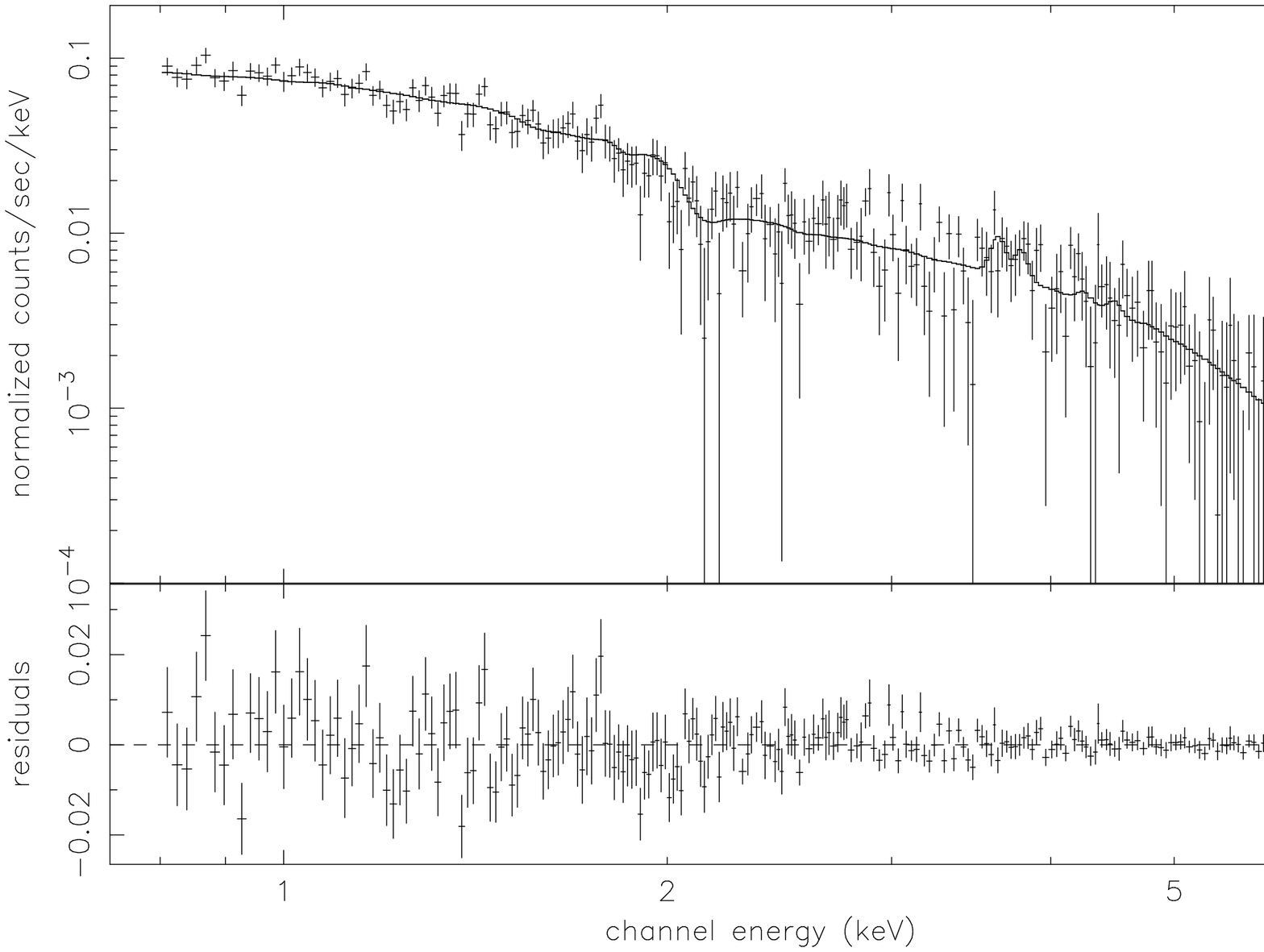}
\figcaption{X-ray spectrum and residuals for MS~1054$-$0321. The spectrum was binned so that there were a minimum of 20 counts per energy bin. {\itshape Solid line:\/} Best-fit model.  \label{fig-1}}
\end{figure}

\placefigure{fig-1}

\begin{figure}
\epsscale{0.6}
\plotone{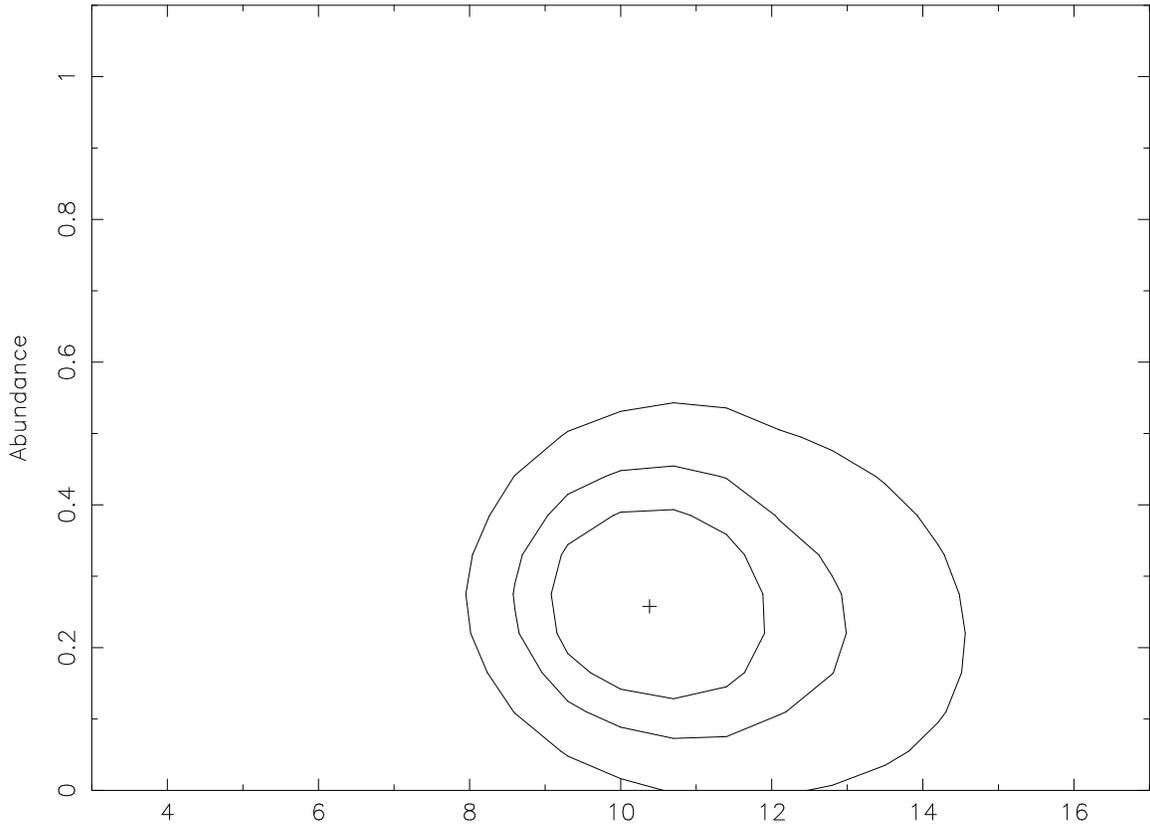}
\figcaption{68.3\%, 90\%, and 99\% confidence $\chi^2$ contours ($\vartriangle\chi^2 = $2.30, 4.61, and 9.21) for the cluster iron abundance and temperature. \label{fig-2}}
\end{figure}

\placefigure{fig-2}

The best fit cluster temperature from the {\itshape Chandra\/} data is somewhat lower than the {\itshape ASCA\/} temperature of $12.3^{+3.1}_{-2.2}$ keV (D98).  However, the two results agree within the 90\% limits, and there is no statistically significant discrepancy.  In addition, the {\itshape ASCA\/} results could be affected by point sources that {\itshape ASCA\/} could not resolve.  For a 2$^{\prime}$ radius region around the cluster including point sources, we find a best fit temperature of $11.4^{+2.5}_{-1.7}$ keV.

We also investigated the variation of temperature with radius by extracting spectra in five concentric annuli.  Annuli were required to have at least 2000 counts above background, which could only be achieved out to a radius of 1.14$^{\prime}$.  We do not find a significant temperature gradient.  The best-fit temperatures vary between annuli by over 2 keV.  However, all temperatures overlapped at the 90\% confidence level.  A plot of temperature as a function of radius is shown in Figure 3.

\begin{figure}
\plotone{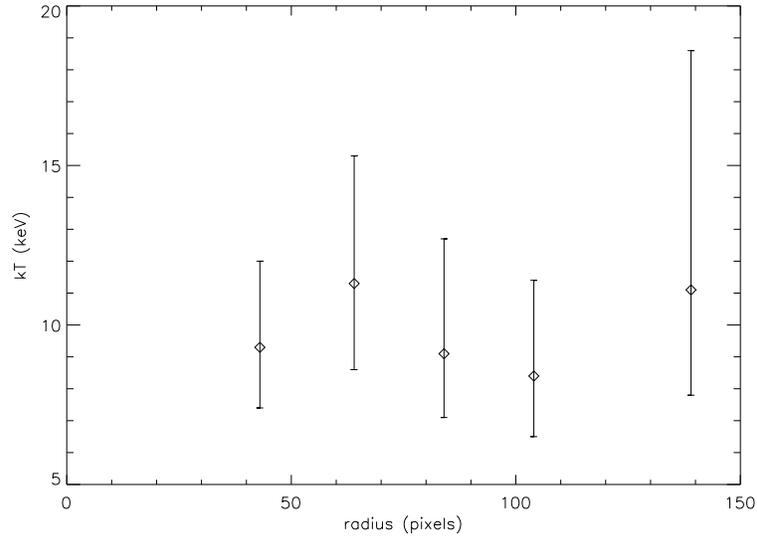}
\figcaption{X-ray temperature as a function of radius.  Temperatures were determined in concentric annuli, and the radius used is the outer radius of each annulus.  The temperature error bars give the 90\% confidence limits.  1 pixel = 2$^{\prime\prime}$.03 = 2.04 $h^{-1}$ kpc. \label{fig-3}}
\end{figure}

\placefigure{fig-3}

\subsection{Substructure}

The {\itshape Chandra\/} image of MS~1054, including the point sources, is shown in Figure 4.  The image has been smoothed with the CIAO program {\itshape csmooth\/} with a minimum significance of 3 and a maximum significance of 5.  The existence of substructure is evident as two distinct peaks can be seen in the X-ray image separated by about $1.2^{\prime}$ (300 $h^{-1}$ kpc).  Figure 5 shows a Hubble Space Telescope WFPC2 I-band mosaic (F814W) of the cluster assembled from the HST archive (for discussion of HST data see van Dokkum et al. 2000) with our X-ray contours overlaid.  In the HST image, MS~1054 appears as an extended string of galaxies.  There are several galaxies centered on the eastern X-ray peak, including the central cD galaxy.  However, the western peak appears to have galaxies lining its southern edge rather than at the center.

We investigated the temperature of each of these clumps separately, taking spectra in a 0.41$^{\prime}$ radius circle around the brightest point in each clump.  The eastern clump was taken to be centered at R.A.(2000) = 10$^h$57$^m$0$^s$.2 and decl.(2000) =  $-$03$^{\circ}$37$^{\prime}$39$^{\prime\prime}$.6, and the western peak center was R.A.(2000) = 10$^h$56$^m$55$^s$.6 and decl.(2000) = $-$03$^{\circ}$37$^{\prime}$42$^{\prime\prime}$.5.  The spectra were again fit with a Raymond-Smith model with the absorbing $N{_H}$ fixed at $3.6 \times 10^{20}$ cm$^{-2}$, the redshift fixed at 0.83, and the iron abundance and temperature free to vary.

\begin{figure}
\plotone{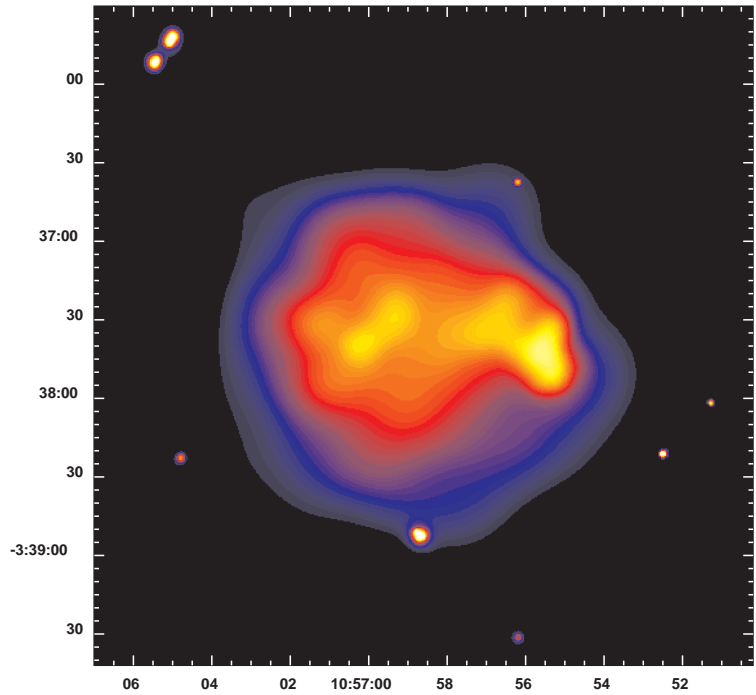}
\figcaption{Smoothed {\itshape Chandra\/} image of MS~1054.  The image was smoothed with the CIAO program {\itshape csmooth\/}. \label{fig-4}}
\end{figure}

\placefigure{fig-4}

\placefigure{fig-5}

For the eastern clump, the best fit temperature is $10.5^{+3.4}_{-2.1}$ keV with a reduced $\chi^2$ of 0.94 for 75 degrees of freedom.  This temperature agrees well with the overall cluster temperature.  For the western clump, the temperature is $6.7^{+1.7}_{-1.2}$ keV with a reduced $\chi^2$ of 1.08 for 62 degrees of freedom.  The eastern clump is somewhat hotter than the western clump, with the lower limit of the 90\% confidence range for the eastern clump just equaling the upper limit for the western clump.  The spectra and best fit folded models are shown in Figure 6.  We find the iron abundances of the eastern and western clumps to be $0.08^{+0.23}_{-0.08}$ and $0.46^{+0.27}_{-0.26}$ respectively, which are consistent with each other and with the value $0.26\pm0.15$ derived for the whole cluster.  Figure 7 shows the abundance versus temperature contours for both clumps.

Within a radius of 0.41$^{\prime}$, the two clumps have similar luminosities.  The eastern clump luminosity is $1.4 \times 10^{44}$ h$^{-2}$ ergs s$^{-1}$ (2-10 keV) and $2.6 \times 10^{44}$ h$^{-2}$ ergs s$^{-1}$ bolometric, and for the western clump the luminosity is $1.0 \times 10^{44}$ h$^{-2}$ ergs s$^{-1}$ (2-10 keV) and $1.7 \times 10^{44}$ h$^{-2}$ ergs s$^{-1}$ bolometric ($q_0 = 0.1$).  The western clump has a higher central surface brightness, but is also more compact.  The eastern clump is comparatively diffuse, whereas the western clump is smaller and denser. 

\placefigure{fig-6}

\placefigure{fig-7}

To get an indication of temperature variations in the cluster, we created a map of the hardness ratio.  First, we made hard (1.5-7 keV) and soft (0.3-1.5) band images, smoothing each using {\itshape csmooth\/} and the output smoothing scales, the kernel size at which the signal to noise is greater than three, from the full band image in Figure 4.  From each image we subtracted the background in the appropriate energy band.  We also discarded pixels with less than twice the background to reduce inaccuracies near the edge of the cluster where we have low counts.  Finally, the hard band was divided by the soft band image.  This hardness ratio map is shown in Figure 8.  Lighter colors correspond to a higher hardness ratio and may indicate higher temperatures.  The full band contours are overlaid on the image.  The eastern clump does appear to have a higher hardness ratio and, therefore, a higher temperature than the western clump.  There does not appear to be a shock between the two clumps.  Indeed, the interclump region is cooler than the cluster as a whole.

\placefigure{fig-8} 

Substructure in MS~1054 was seen previously in the {\itshape ROSAT/HRI\/} observation, which has lower resolution than {\itshape Chandra\/} (NA00, D98).  NA00 also find two components and identify the eastern clump as the main cluster component.  Indeed our position of the eastern clump is quite close to the central cD galaxy which has R.A.(2000) = 10$^h$56$^m$59$^s$.9 and decl.(2000) =  $-$03$^{\circ}$37$^{\prime}$37$^{\prime\prime}$.3 (D98).  The western clump could either be a subgroup falling into the cluster or a foreground group of galaxies.  Since we detect an iron line, we fit for the redshift of the western clump in XSPEC and get a value of z = $0.84\pm0.02$, which agrees well with the redshift of the cluster.

Substructure can also be seen in the weak lensing analysis of the optical data.  In their weak lensing study, Hoekstra et al. (2000) find three clumps in the mass distribution that all appear to have similar masses.  An approximate overlay of the Chandra X-ray contours on their weak lensing mass reconstruction is shown in Figure 9.  Their central and western clumps seem to correspond to our eastern and western clumps.  However, we do not detect the north-eastern weak lensing clump.  This result was noted by Clowe et al. (2000) when they overlaid the {\itshape ROSAT/HRI\/} contours on the Keck weak lensing data.  It is possible that this third peak is not as fully collapsed as the other two and, therefore, is not yet visible in X-rays.  Alternatively, reanalysis of the weak lensing data shows that the details of the substructure are not well constrained by the data (Marshall 2001).

\placefigure{fig-9}

We used the galaxy redshifts published in Tran et al. (1999) to estimate the mean velocity of each of the three weak lensing clumps.  For the north-eastern, central, and western clumps the mean velocities and one sigma errors are $161,700\pm280$ km s$^{-1}$, $162,600\pm410$ km s$^{-1}$, and $162,300\pm350$ km s$^{-1}$ respectively.  The difference in the velocities of the unseen north-eastern peak and central peak is $900\pm690$ km s$^{-1}$, indicating that a significant relative velocity is possible but not required by the limited data (6-7 galaxies per clump).

In the HST image, MS~1054 has a filamentary appearance.  It is possible that this cluster is in reality a series of clumps along a filament that are in the process of merging.  The clear separation of the two X-ray peaks indicate that the cluster is most likely in a pre-merger state.  When considering the overlay of the X-ray contours on both the Hubble mosaic and the weak lensing mass map it appears that both the cluster galaxies and mass associated with the western subclump lie below the X-ray peak.  This leads to the interesting possibility that the gas in the subclump is being stripped off as it falls into the cluster.

In light of this cluster's complicated structure using a spherical, hydrostatic, isothermal model may lead to errors in estimating the mass.  To get an indication of these errors, we turn to cluster simulations.  Simulations show that mergers can introduce large errors in mass estimates using X-ray properties, but generally produce underestimates (Mathiesen \& Evrard 2001; Evrard, Metzler, \& Navarro 1996; Roettiger, Burns, \& Loken 1996; Schindler 1996).  In section 4 we estimate the virial mass of the cluster using a scaling relation between the mass and X-ray temperature.  Scaling relations between cluster virial mass and temperature are normalized through cluster simulations and use either a mass weighted or emission weighted temperature.  These temperatures are not necessarily the same as the spectral temperature obtained from observations.  

Recently, Mathiesen \& Evrard (2001) have used simulations to investigate the relationship between these three temperatures.  As they note, one might expect that a subclump containing cooler gas would cause an underestimation of the cluster temperature and mass, whereas a shock containing hotter gas would cause an overestimate of these properties.  However, the larger effect will be due to the subclump which is more massive and luminous than the gas in the shock, and mergers will lead to an underestimate of the mass.  Indeed, they find that the spectrally determined temperatures are nearly always lower than both the emission and mass-weighted temperatures with typical errors of 10-20\%.  In the case of the mass-weighted temperature they show that all large underestimates of the temperature occur close to a merger.  We conclude that if MS~1054 is not in hydrostatic equilibrium, then we should underestimate its mass.  This only strengthens our conclusions about $\Omega_m$.

\subsection{Point Sources}

We resolve nine point sources within a 3$^{\prime}$ radius of the cluster center.  From a NASA/IPAC Extragalactic Database (NED) search, one of these corresponds to LCRS~B105416.2$-$032123, which is an AGN detected at radio frequencies with a redshift of 0.2.  Our position for this source is R.A.(2000) = 10$^h$56$^m$48$^s$.7 and decl.(2000) = $-$03$^{\circ}$37$^{\prime}$27$^{\prime\prime}$.4, and the NED quoted position is R.A.(2000) = 10$^h$56$^m$48$^s$.8 and decl.(2000) = $-$03$^{\circ}$37$^{\prime}$26$^{\prime\prime}$.  One of the other sources is in the vicinity of one of the MS~1054 galaxies used by D98 in studying the velocity dispersion, R.A.(2000) = 10$^h$56$^m$52$^s$.5 and decl.(2000) = $-$03$^{\circ}$38$^{\prime}$21$^{\prime\prime}$.5 compared to NED of R.A.(2000) = 10$^h$56$^m$53$^s$.3 and decl.(2000) = $-$03$^{\circ}$38$^{\prime}$16$^{\prime\prime}$.  The remaining sources do not seem to have any obvious NED couterparts.

Chapman et al. (2000) report a SCUBA source, SMMJ~10571-0337, near MS~1054.  This source has a 850 $\mu$m flux of 15mJy, but is not detected by {\itshape Chandra\/}.  Based on our non-detection of this source, we estimate the upper limit on its flux in the 2-7 keV range to be $1.3 \times 10^{-15}$ ergs cm$^{-2}$ s$^{-1}$.

The point source closest to the cluster is centered at R.A.(2000) = 10$^h$56$^m$58$^s$.7 and decl.(2000) = $-$03$^{\circ}$38$^{\prime}$53$^{\prime\prime}$.5 and was seen in the {\itshape ROSAT\/} image as a small southern extention to the cluster.  Fitting to an absorbed power law, gives $N_H = 2.2^{+1.8}_{-1.4} \times 10^{21}$ atoms/cm$^2$, and photon index 1.7$\pm0.3$, with a reduced $\chi^2$ of 1.26 for 20 degrees of freedom.  This photon index is typical of a Seyfert I type AGN.  The binned spectrum and folded model are shown in Figure 10.  The flux of this source is approximately $3.1 \times 10^{-14}$ ergs cm$^{-2}$ s$^{-1}$ (2-7 keV).  Although this object was not found in the NED database, there is a small object in the HST image that may correspond to it.

\placefigure{fig-10}

\section{CLUSTER MASS AND VELOCITY DISPERSION}

Using the X-ray temperature for the full cluster, $10.4^{+1.7}_{-1.5}$ keV, we can estimate both the velocity dispersion and mass of MS~1054 and compare to optical results.  From the {\itshape Chandra\/} temperature the implied velocity dispersion is $(kT_X/$$\mu$$m_p)^{1/2} = 1289^{+102}_{-96}$ km s$^{-1}$.  This agrees well with the observed velocity dispersion of $1170\pm150$ km s$^{-1}$ (Tran et al. 1999) and the velocity dispersion derived from weak lensing of $1311^{+83}_{-89}$ km s$^{-1}$ (Hoekstra, et al. 2000).  

The virial mass of the cluster can be estimated for $\Omega_m$ = 1 by assuming that the mean density in the virialized region is $\sim$200 times the critical density at the cluster's redshift and that the cluster is isothermal (Evrard et al. 1996, Hjorth, Oukbir, \& van Kampen 1998, D98).  From the simulations of Evrard et al. (1996), the mass-temperature relation for the mass within a region whose density is 200 times the critical density is
\begin{equation}
M_{vir} \approx (1.45 \times 10^{15} h^{-1} M_{\odot})\biggl(\frac{1}{1+z}\biggr)^{3/2}\biggl(\frac{kT_X}{10 keV}\biggr)^{3/2}
\end{equation}
(Arnaud \& Evrard 1999).  This scaling law method of estimating the mass of clusters is more accurate than estimates derived from a $\beta$-model fit to the surface brightness because the scaling law has a smaller variance and is less sensitive to the cluster's dynamical state.  However, the scaling law method depends on the normalization of the M-T relation which is derived from cluster simulations (Evrard et al. 1996).  Bryan \& Norman (1998) compare the normalizations from several cluster simulation studies and find that the scatter in normalizations is small, similar to the scatter of clusters around the M-T relation.  We use the relation given by Evrard et al. (1996) because it gives the smallest mass.  As we noted before, we expect to underestimate the mass due to the presence of substructure and this can only strengthen our conclusions in the following section.

From the {\itshape Chandra\/} temperature, the virial mass is approximately $6.2^{+1.6}_{-1.3} \times 10^{14} h^{-1} M_{\odot}$ within $r_{200} = 0.76 h^{-1}$ Mpc.  The errors here represent the errors in the temperature.  This mass is somewhat lower than those derived from weak lensing, $M(\leq 0.87 h_{50}^{-1} Mpc) = 1.2\pm0.2 \times 10^{15}h_{50}^{-1}M_{\odot}$, and the observed velocity dispersion, $M(\leq 1 h^{-1} Mpc) = 1.9\pm0.5 \times 10^{15}h^{-1}M_{\odot}$ (Hoekstra, Franx, \& Kuijken 2000; Tran et al. 1999).  Assuming $M \propto R$ in order to compare with our mass, these masses are approximately $M(\leq 0.76 h^{-1} Mpc) = 1.0\pm0.2 \times 10^{15}h^{-1}M_{\odot}$ and $M(\leq 0.76 h^{-1} Mpc) = 1.4\pm0.4 \times 10^{15}h^{-1}M_{\odot}$ respectively.

Another method of estimating the mass comes from fitting a beta-model to the observed surface brightness.  The radial surface brightness profile is modelled as $S = S_0(1+r^2/r_c^2)^{-3\beta+0.5}$ where $r_c$ is the core radius.  The background subtracted radial surface brightness profile is shown in Figure 11.  Attempts to fit to a $\beta$-model failed because the best fit was obtained for unreasonably large values of $\beta$ and the core radius.  This behavior was also noted in NA00 for the {\itshape ROSAT/HRI\/} data.  NA00 found that if they removed the western substructure they could get a good fit to a $\beta$-model; however, we were not able to get a reasonable fit even with the western clump removed.  In their simulated clusters, Bartelmann \& Stienmetz (1996) find that they get more accurate masses by constraining $\beta$ to be one and fitting for the core radius, though their masses are still biased low by about 10\%.  Following this method, we get a best-fit core radius of 1.1$^{\prime}$ (270 kpc).  This core radius is rather large, on the order of the size of the cluster.  However, using Eq. 22 from Bartelmann \& Stienmetz (1996), we get a mass of $M(\leq 0.76 h^{-1} Mpc) = 7.4 \times 10^{14}h^{-1}M_{\odot}$.  This agrees well with the mass we derive from the M-T relation. 

\placefigure{fig-11}

\section{CONSTRAINTS ON $\Omega_m$}

Following a line of reasoning similar to that used in D98, we determine the expected number density of clusters like MS~1054 in an $\Omega_m$ = 1 universe with initial Gaussian perturbations and compare this with the observed number density for detection in the EMSS.  The comoving mass density of virialized objects with masses greater than $M$ is given by the Press-Schechter formula
\begin{equation}
\rho(>M) = \rho_0\/ erfc(\frac{\nu_c}{\sqrt{2}}) = \frac{2\rho_0}{\sqrt{\pi}}\int_{\nu_c/\sqrt{2}}^{\infty}e^{-x^2}\,dx,
\end{equation}
where $\rho_0$ is the current matter density, and $\nu_c$ is the critical threshold at which perturbations virialize (Press \& Schechter 1974, D98).

We conservatively take the temperature of MS~1054 to be greater than 8.5 keV; D98 used 10 keV for this calculation.  The cluster mass is then at least $4.5 \times 10^{14} h^{-1} M_{\odot}$.  The current number density of clusters of a given mass has been estimated from both observations and simulations (Bahcall \& Cen 1992, Eke et al. 1996, Bahcall et al. 1997).  We will use the results of Bahcall et al. (1997) as they give us the highest number density.  We, therefore, find that at $z = 0$ the number density is less than $1.1 \times 10^{-6} h^3$ Mpc$^{-3}$.  The mass density is then approximately $\rho(>M) \sim nM \sim 4.9 \times 10^8 h^2 M_{\odot}$ Mpc$^{-3}$.

From the Press-Schechter formula with this mass density, $\nu_c(z=0)$ is greater than 3.13.  If $\Omega_m$ = 1, $\nu_c$ is proportional to (1+z), and $\nu_c(z=0.83)$ is greater than 5.72 for clusters with $T_X > 8.5$ keV.  This corresponds to a mean virialized mass density of like clusters at z=0.83 less than $2970 h^2 M_{\odot}$ Mpc$^{-3}$, and a number density less than $6.6 \times 10^{-12} h^3$ Mpc$^{-3}$.  However, detection of MS~1054 in the EMSS gives a density of like clusters of $\sim 10^{-8} h^3$ Mpc$^{-3}$ (D98).  This exceeds our prediction for an $\Omega_m$ = 1 universe with Gaussian perturbations by a factor of more than $10^3$.  Although the lower {\itshape Chandra\/} temperature gives an expected number density a factor of 15 higher than that found in D98, it does not change their conclusion that $\Omega_m$ = 1 is very unlikely.

Donahue and Voit (1999) fit for $\Omega_m$ using three cluster samples with different redshift ranges.  They find $\Omega_m \simeq 0.45$ for an open universe and $\Omega_m \simeq 0.27$ for a flat universe.  These results are unaffected by the lower temperature for MS~1054.  In fact, with a temperature of 10 keV, MS~1054 appears to lie closer to the best fit temperature function for the high redshift sample.

\section{SUMMARY}

The {\itshape Chandra\/} observation of the cluster MS~1054$-$0321 at z = 0.83 indicates that it has an X-ray temperature of $10.4^{+1.7}_{-1.5}$ keV.  This is somewhat lower than, but consistent with, the temperature of $12.3^{+3.1}_{-2.2}$ keV previously found from the {\itshape ASCA\/} data (D98).  We are also able to make the first positive identification of an iron line in this cluster and find a value of $0.26\pm0.15$ for the abundance relative to solar (Anders \& Grevessa 1989).  The detection of iron in a cluster at this redshift is consistent with early enrichment of the ICM (Mushotzky \& Loewenstein 1997).  {\itshape Chandra\/} was able to resolve substructure in MS~1054, confirming the substructure in the {\itshape ROSAT/HRI\/} observation, and to identify a number of point sources surrounding it, a task beyond the spatial resolution of the {\itshape ROSAT/HRI\/}.  The X-ray distribution appears to have two clumps.  For the eastern clump, probably associated with the main cluster component, we find an X-ray temperature of $10.5^{+3.4}_{-2.1}$ keV.  For the western clump we find a slightly lower temperature of $6.7^{+1.7}_{-1.2}$ keV.  The best-fit redshift of this second clump corresponds to the cluster redshift, and it may be a subgroup falling into the cluster. The hardness ratio map shows no evidence of a shock between the two clumps; however, comparison of the X-ray contours with the Hubble image and weak lensing mass reconstruction may indicate that the gas in the subclump is being stripped off as it falls into the cluster.  We do not detect a third clump seen in the weak lensing derived mass distribution.

In this work, we confirm, along with the weak lensing and velocity dispersion data, that MS~1054 is truly a massive cluster.  With {\itshape Chandra\/}'s resolution, we are able for the first time to determine the temperature of different regions of the cluster.  This allows us to confirm that the entire cluster is hot, and that we are not measuring an anomalously high temperature due to a small shock-heated region.  With {\itshape Chandra\/}, we are also able to quantify the point source contamination and remove it from the data when determining the temperature.

The velocity dispersion derived from our temperature is in good agreement with the observed velocity dispersion and weak lensing estimates, although our mass estimate is somewhat lower than the masses derived from these methods (Tran et al. 1999; Hoekstra et al. 2000).  The lower X-ray temperature from {\itshape Chandra\/} of course leads to a smaller derived virial mass than the {\itshape ASCA\/} estimate.  Even with this smaller mass, the predicted number density of clusters like MS~1054 assuming Gaussian perturbations in an $\Omega_m = 1$ universe is much smaller than the observed density.  Cluster simulations indicate that we may underestimate the cluster's mass owing to the presence of substructure, but this does not affect our conclusions.  Given the assumption of gaussian initial fluctuations, this cluster still provides convincing evidence for $\Omega_m < 1$.  When seen in conjunction with cosmic microwave background experiments that find $\Omega_{tot} \sim 1$ (Miller et al. 1999; de Bernardis et al. 2000, Hanany et al. 2000), this implies the existence of dark energy in the universe independant of the supernova result (Riess et al. 1998; Perlmutter 1999).

\acknowledgments

We would like thank Greg Bryan for the very helpful discussion of cluster simulations and mass estimates.  We would also like to thank Herman Marshall, Mike Wise, and the rest of the MIT HETG/CXC group.  Special thanks go to Megan Sosey  of the Space Telescope Science Institute for her work on the Hubble mosaic image.  This work was funded by contract SAO SV1-61010, NASA contracts NAS8-39073, NAS8-37716, and NAS8-38252, and HST grant GO-6668.  MED was also partially supported by NAG5-3257 and NAG5-6236.  TEJ would like to acknowledge the support of a NSF fellowship.

%starting of references

\clearpage

%figures

\begin{figure}
%\plotone{f5.eps}
\figcaption{Hubble I-band mosaic of MS~1054 with contours from the smoothed X-ray image overlaid.  Contours are evenly spaced from 0 to 1 by 0.10.  \label{fig-5}}
\end{figure}

\begin{figure}
\epsscale{0.7}
\plottwo{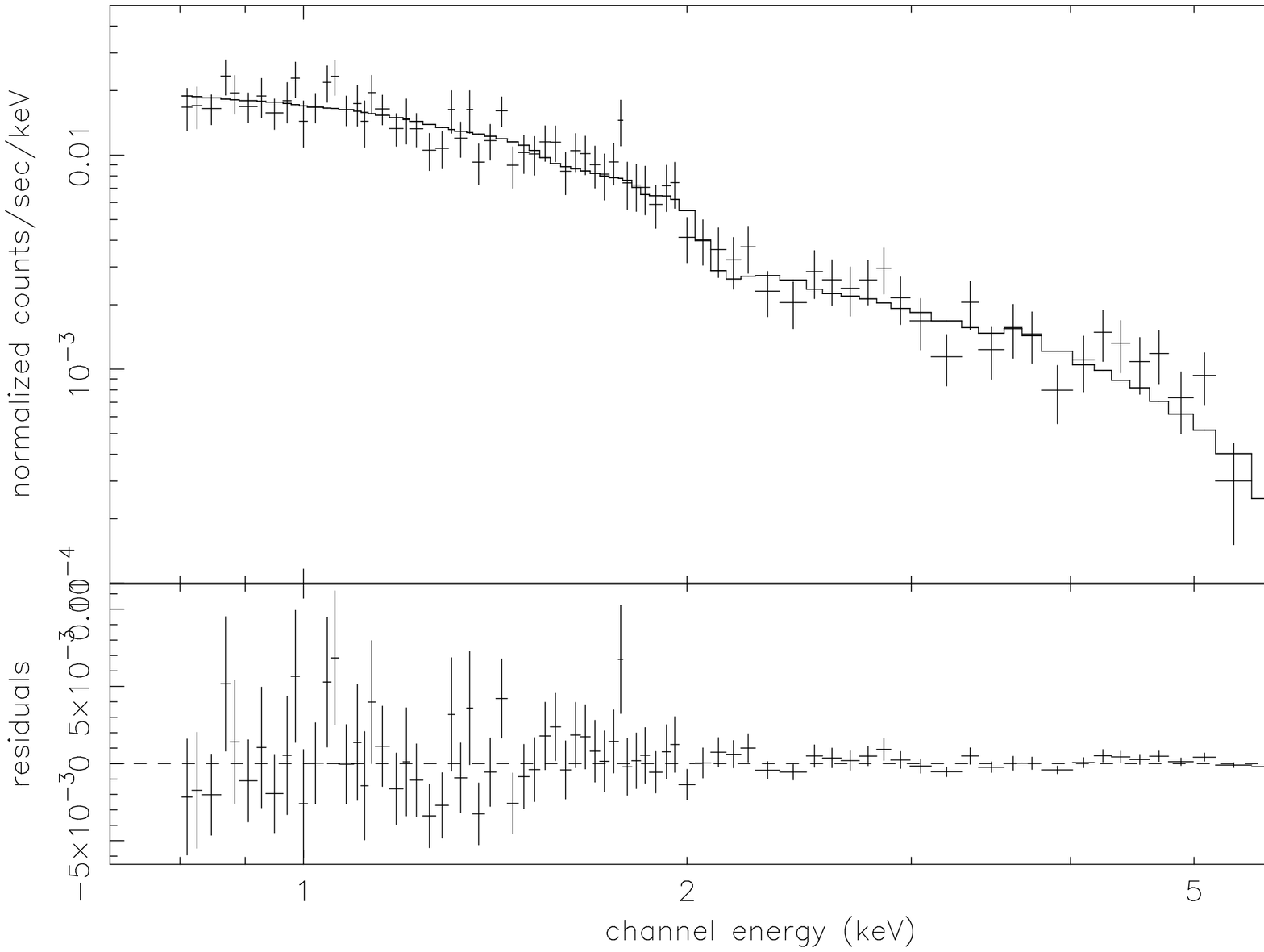}{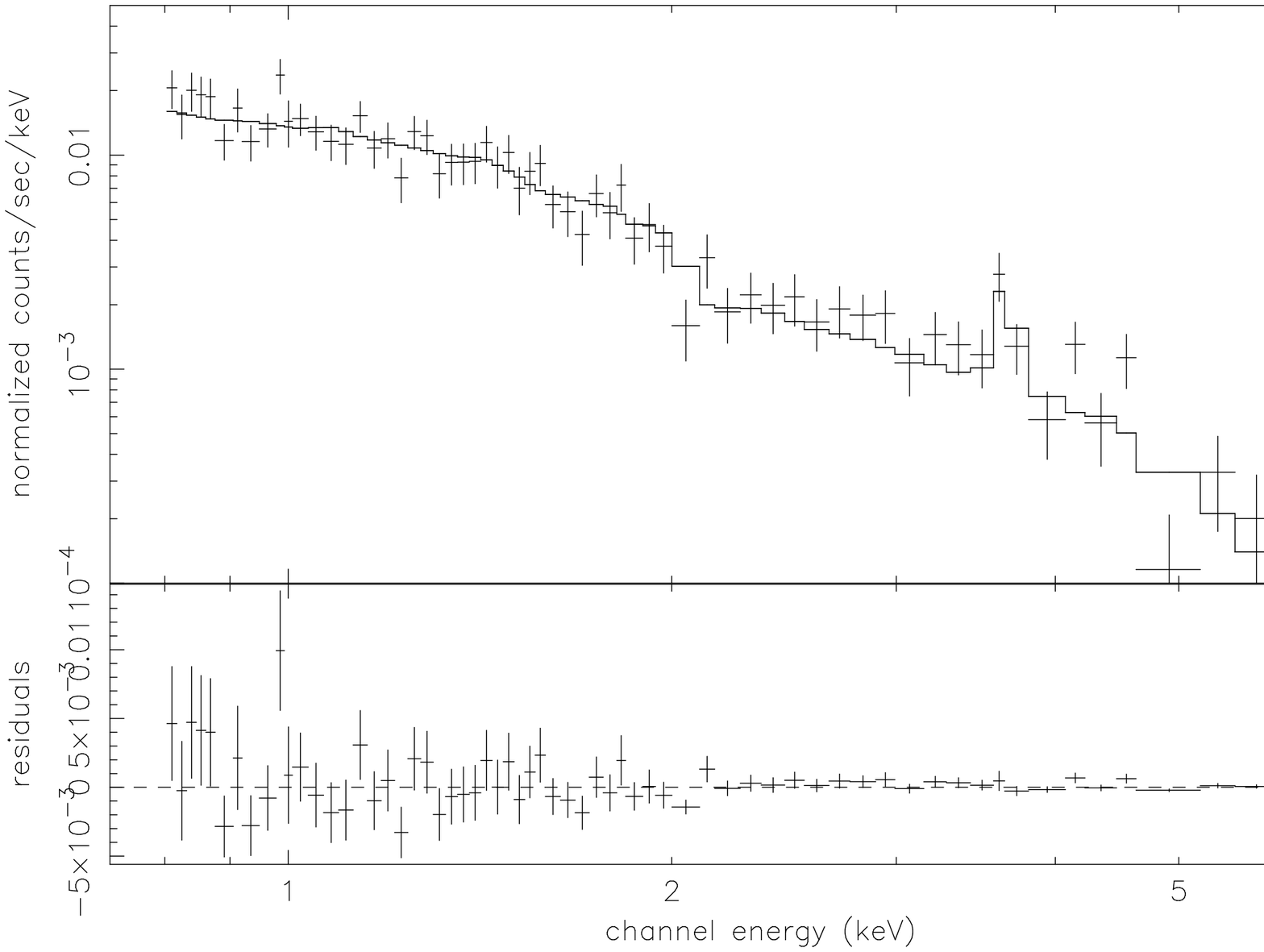}
\figcaption{X-ray spectrum and residuals for a 0.41$^{\prime}$ radius circle around the (a) eastern peak and (b) western peak. {\itshape Solid line:\/} Best-fit model.  \label{fig-6}}
\end{figure}

\begin{figure}
\epsscale{0.6}
\plotone{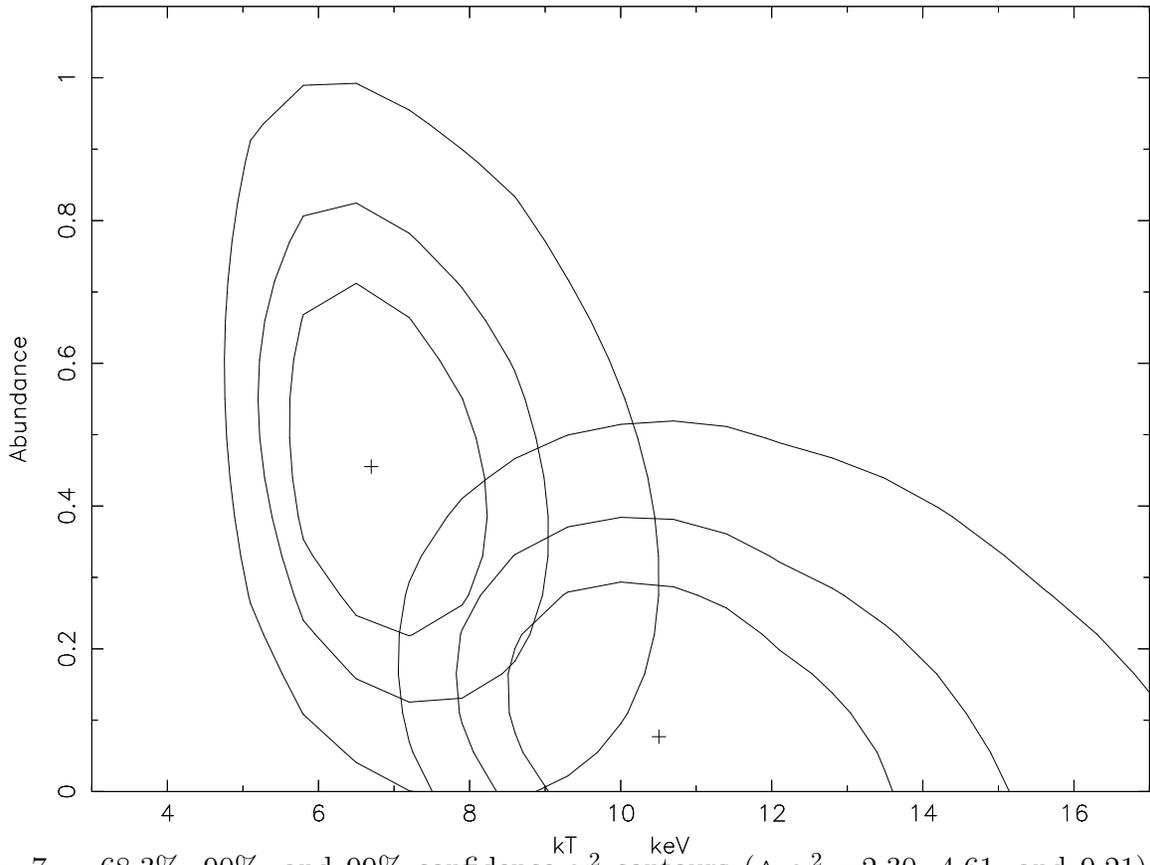}
\figcaption{68.3\%, 90\%, and 99\% confidence $\chi^2$ contours ($\vartriangle\chi^2 = $2.30, 4.61, and 9.21) for the iron abundance and temperature of the eastern and western peaks. The eastern peak contours are centered at kT$ = 10.5$ keV and an abundance of $0.08$.  The western peak contours are centered at kT$ = 6.7$ keV and an abundance of $0.46$. \label{fig-7}}
\end{figure}

\begin{figure}
\plotone{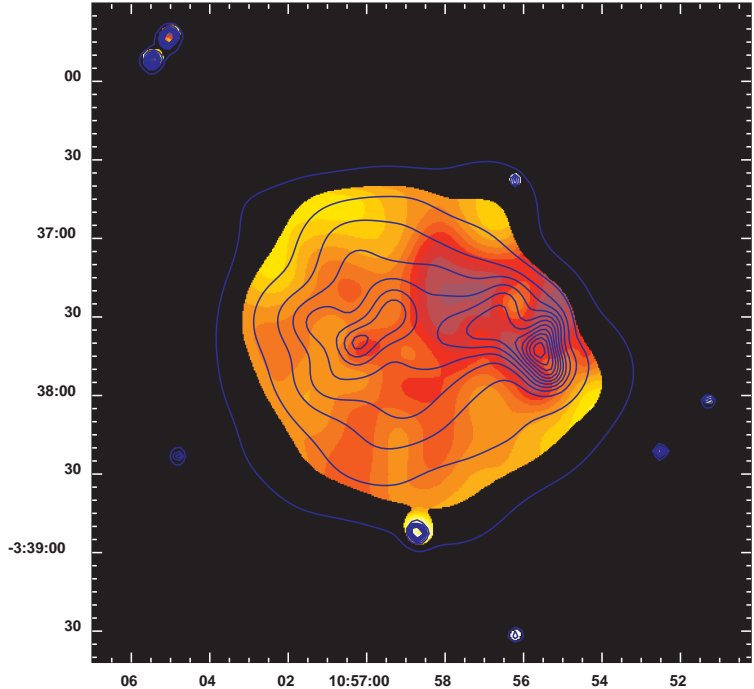}
\figcaption{Hardness ratio map with full band contours overlaid.  The smoothed hard band (1.5-7 keV) image was divided by the smoothed soft band (0.3-1.5 keV) image after taking out pixels with less than twice the background.  Lighter colors correspond to a higher ratio.  Contours are evenly spaced from 0 to 1 by 0.067.  \label{fig-8}}
\end{figure}

\begin{figure}
%\epsscale{1.0}
%\plotone{f9.eps}
\figcaption{Weak lensing mass reconstruction from Hoekstra et al. (2000) overlaid with X-ray contours.  The position of the contours with respect to the mass profile is approximated, but relative postions are good to within 3$^{\prime\prime}$.  The origin corresponds to the central cD galaxy, and negative x values indicate east of this origin.  \label{fig-9}}
\end{figure}

\begin{figure}
\epsscale{0.6}
\plotone{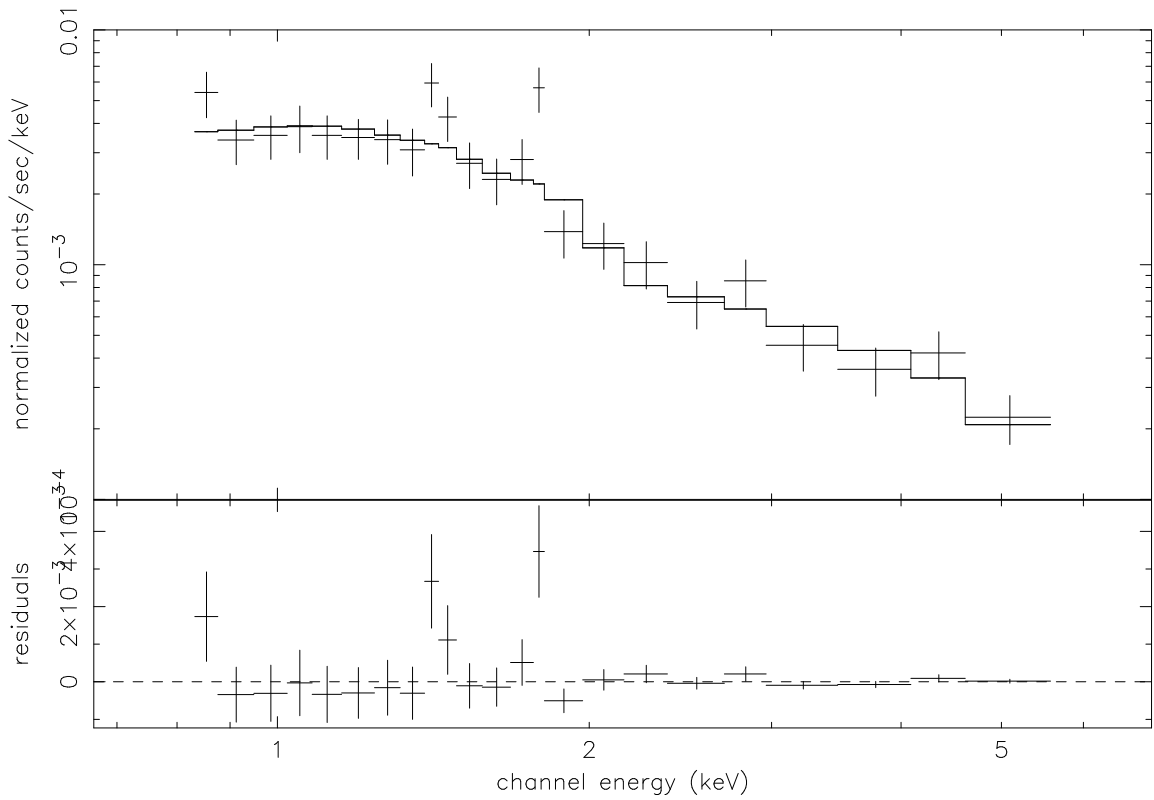}
\figcaption{X-ray spectrum and residuals for the point source centered at R.A.(2000) = 10$^h$56$^m$58$^s$.7 and decl.(2000) = $-$03$^{\circ}$38$^{\prime}$53$^{\prime\prime}$.5. The spectrum was binned so that there were a minimum of 20 counts per energy bin. {\itshape Solid line:\/} Best-fit model.  \label{fig-10}}
\end{figure}

\begin{figure}
\plotone{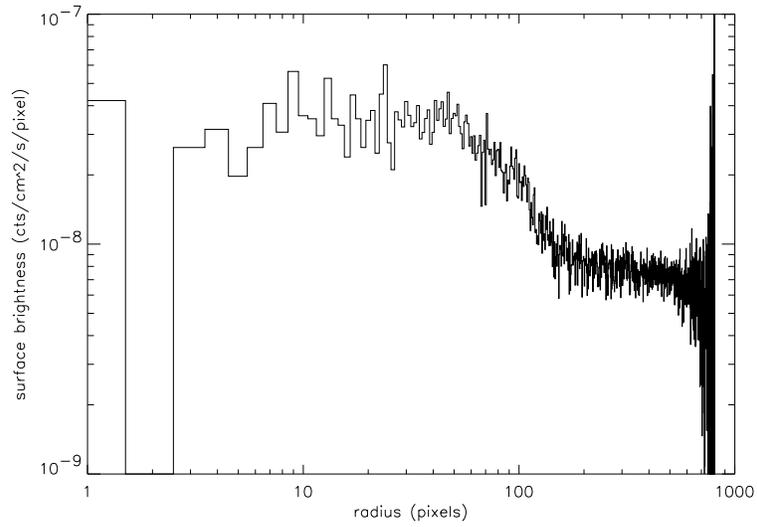}
\figcaption{Radial surface brightness profile for MS~1054$-$0321.  1 pixel = 2$^{\prime\prime}$.03 = 2.04 $h^{-1}$ kpc.  \label{fig-11}}
\end{figure}

\end{document}